# Correlation between local structure and dynamic heterogeneity in a metallic glass-forming liquid

S. P. Pan[a,b,*], S. D. Feng[c], J. W. Qiao[a,b], W. M. Wang[d], and J. Y. Qin[d]

[a]*College of Materials Science and Engineering, Taiyuan University of Technology, Taiyuan, 030024, China*

[b]*Shanxi key laboratory of advanced magnesium-based materials, Taiyuan University of Technology, Taiyuan, 030024, China*

[c]*State Key Laboratory of Metastable Materials Science and Technology, Yanshan University, Qinhuangdao 066004, China*

[d]*Key Laboratory for Liquid-Solid Structural Evolution and Processing of Materials (Ministry of Education), Shandong University, Jinan 250061, China*

**Abstract**

Dynamic heterogeneity as one of the most important properties in supercooled liquids has been found for several decades. However, its structural origin remains open for many systems. Here, we propose a new structural parameter to characterize local atomic packing in metallic liquids. It is found that the new parameter in a simulated metallic glass-forming liquid is closely correlated with potential energy and atomic mobility. It also exhibits significant spatial heterogeneities and these structural fluctuations show close correlation with the spatial distribution of the long-time dynamic propensities. Therefore, our results provide a direct evidence of the correlation between atomic structure and dynamical heterogeneity

\* Corresponding author.
*E-mail address*: shaopengpan@163.com.

# 1. Introduction

Dynamic heterogeneity is considered as one of the three key physical features which dominate much of the behavior of liquids as they are supercooled (The other two are the massive increase in the shear viscosity and the temperature dependence of the entropy) [1-3]. Although the existence of dynamic heterogeneity has been established for several decades [2-5], an obvious question, what cause it, still has no universal answer. While there has been some significant progress [6-12], the structural origin of dynamic heterogeneity remains open for many systems. The atoms in icosahedral clusters have been proved to move slower than other atoms in some metallic glass-forming liquids [6-7]. However, in some other systems icosahedral clusters are absent [13]. Even in the systems with a large number of icosahedral clusters in the glassy state, the number of icosahedral clusters is rather small at high temperatures in the liquid state [14]. Therefore, the icosahedrons cannot be adopted as a universal structural indicator for dynamic heterogeneity. Thus some other indicators are used to predict dynamic heterogeneity. The Debye-Waller factor has been quite successful for predicting the relative long-time dynamical heterogeneity and irreversible arrangement in glass-forming liquids [15]. The localized soft modes are appreciated to play the central role in the dynamic heterogeneity [16-18]. However, both of the two indicators cannot provide a clear picture of local atomic structure. A more general structural parameter is needed to characterize structural heterogeneity and predict dynamic heterogeneity and even other properties in metallic liquids and glasses.

It is natural to relate dynamic heterogeneity with local free volume [19-22]. However, it is found that local free volume does not show strong correlation with local mobility [23-24]. Instead, Ediger and Harrowell suggested that the dynamic heterogeneity in liquids might be caused by some "defects" which are similar to those in the solid state [1]. However, the structural origin of the "defects" is still unknown. It is well known that the atomic packing is usually much looser around the defects in the crystal. Therefore, the "defects" in liquids might be also located at the regions with loose atomic packing. However, the present structure parameters cannot describe the atomic packing well.

We propose a new local structure parameter, quasi-nearest atom (QNA), to describe the atomic packing in metallic liquids and study the correlation between structural heterogeneity and dynamic heterogeneity in a simulated metallic glass-forming liquid ($Cu_{64}Zr_{36}$). We find that atoms with more QNAs have higher potential energy and higher mobility. The atoms with different number of QNAs show clear spatial heterogeneities and the more QNA regions overlap the regions with higher dynamic propensities, and visa inverse. Therefore, our results provide a direct evidence of the correlation between structural heterogeneity and dynamic heterogeneity in the metallic glass-forming liquid.

## 2. Methods

Classical molecular dynamics simulations are carried out on $Cu_{64}Zr_{36}$ metallic liquid. Our system consists of 128000 Cu atoms and 72000 Zr atoms. The atoms interact via embedded-atom method (EAM) potential [25]. The system is simulated with periodic boundary conditions. Isothermal-isobaric (NPT)-ensemble simulations using the Nose-Hoover thermostat and barostat are employed in our studies. The equations of motion are integrated using the Verlet algorithm with a time step of 1 fs. Eight independent runs are performed for the measurements of structural and dynamical properties at $T=1000K$ and zero external pressure.

The local structure is characterized by QNA, which is defined as follows. All the nearest neighbors around each atom in the system can be determined by Voronoi tessellation method [26]. According to this method, each nearest neighbor of the center atom corresponds to one face of the Voronoi polyhedron. If two Voronoi faces share an edge, the two corresponding atoms are defined as an adjacent pair of atoms. Next, if an adjacent pair of atoms are not the nearest neighbors of each other, we identify these two atoms as a pair of QNAs. For instance, as shown in Figure 1(a), atoms *C* and *D* are a pair of QNAs because they satisfy the following conditions: (i) both of them are the nearest neighbors of atom *A*; (ii) they are adjacent among all the nearest neighbors of atom *A*; (iii) they are not the nearest neighbors of each other. In Figure 1, atom *A* has less QNA than atom *J* while the atomic packing around *A* is much denser than that around *J*. Therefore, QNA can reflect local constraints. The ability to move for an atom is associated with the degree to which the atom is constrained by its surroundings [15]. Thus there might be some correlation between QNA and atomic mobility.

### 3. Results and discussion

We present the distribution of the number of QNAs, $N_Q$, at 1000K in Figure 1(b). It follows similar behavior for both Cu and Zr atoms. However, the fraction of Cu atoms with small $N_Q$ is slightly larger than that of Zr, and the situation is opposite for the atoms with large $N_Q$. This indicates that the local packing of Cu is denser than that of Zr. We note that the fraction shows a peak at $N_Q \sim 2$, suggesting that most of the atoms are not closely packed at 1000K. The distribution of atomic potential energy with different $N_Q$ for Cu and Zr is shown in Figure 1(c) and (d), respectively. It can be seen that the distributions have large overlaps, indicating the correlation between $N_Q$ and atomic potential energy is not a one-to-one correspondence. Considered that the cutoff distance of potential energy is 6.5Å, much larger that the scale of $N_Q$, it is reasonable for the large overlaps. However, as shown in the insets of Figure 1(c) and (d), atoms with larger $N_Q$ have larger per-atom potential energy shown. In this respect, $N_Q$ plays an key role in the correlation between local structure and potential energy. This fact suggests that atoms with larger $N_Q$ tend to have lower thermodynamic stability and thus might move faster.

We next study the structural relaxation with different local structures. We label all the atoms with different $N_Q$ at initial time. We obtain the structural relaxation time for atoms with the same $N_Q$ by calculating the self-intermediate scattering function (SISF) [27],

$$F_s^{ab}(q,t) = \frac{1}{N_{ab}} \sum_{j=1}^{N_{ab}} \left\langle \exp\left\{i\vec{q} \cdot \left[\vec{r}_j(t) - \vec{r}_j(0)\right]\right\}\right\rangle, \tag{1}$$

where $N_{ab}$ is the number of type *a* (either Cu or Zr) atoms with $N_Q = b$ at $t = 0$, $\vec{r}$ is the position of each atom, $\vec{q}$ is the wave vector which corresponds to the first peak of the partial structure factor (2.8 Å$^{-1}$ for Cu and 2.7 Å$^{-1}$ for Zr) and the average is taken over 8 independent runs. Figure 2(a) and 2(b) display the SISFs of Cu and Zr atoms with different $N_Q$. In the long-time relaxation (often called α-relaxation) regime, the SISF with small $N_Q$ decays more slowly compared to that with larger $N_Q$. This indicates that atoms with smaller $N_Q$ tend to move slower than those with larger $N_Q$. The α-relaxation time is defined as the time at which the SISF decays to 1/$e$ of its initial value. As shown in the insets of Figure 2(a) and 2(b), for either component, the relaxation time for atoms decreases with increasing $N_Q$.

The above results clearly show a correlation between local structure and dynamics, yet the dynamic behavior of an individual atom is missing. We further quantify the dynamic heterogeneity using an alternative method, to calculate the dynamic propensity of an atom, $\langle \Delta r_i^2 \rangle_{ic}$, which is defined as $\langle [\vec{r}_i(t) - \vec{r}_i(0)]^2 \rangle$, where the average is taken over the ensemble of $N$-particle trajectories, all starting from the same configuration but with momenta assigned randomly from the appropriate Maxwell-Boltzmann distribution [28]. The time interval needs to be chosen to maximize the observed dynamic heterogeneities. Here we have chosen the interval to be the "maximum non-Gaussian time", at which the non-Gaussian parameter, $\alpha_2(t) = 3<r^4(t)>/5<r^2(t)>^2 - 1$, reaches the maximum. In this work, the "maximum non-Gaussian time" is about 3.5 ps at 1000K, which is longer than two times of the relaxation time of the system. In Figure 2(c) and 2(d), we plot the distribution of dynamic propensities for Cu and Zr in 8 independent configurations at 1000K, averaging over 100 runs. For either component, the distribution of propensities for atoms with $N_Q = 0$ has the highest peak at the end of low propensities, which indicates most of the atoms with $N_Q = 0$ tend to move slowly. As $N_Q$ increases, the peak becomes a little lower, and move to the higher propensities. In the insets we show that the average propensity increases with $N_Q$. All these show a close correlation between $N_Q$ and atomic mobility although this correlation is not a one-to-one correspondence.

To quality the correlation between $N_Q$ and $\langle \Delta r_i^2 \rangle_{ic}$, the atoms are sorted by their $\langle \Delta r_i^2 \rangle_{ic}$ from low to high for each element. Then they are divided into $n_g$ groups, each containing $n_A$ atoms. For each group, the average $N_Q$ ($<N_Q>$) and $\langle \Delta r_i^2 \rangle_{ic}$ ($\langle\langle \Delta r_i^2 \rangle_{ic}\rangle$) are calculated. Figure 3(a) shows $<N_Q>$ exhibits a linear relation with $\langle\langle \Delta r_i^2 \rangle_{ic}\rangle$ in log-semi plot with $n_g$ = 200 For Cu and Zr, indicating an exponential dependence of $\langle\langle \Delta r_i^2 \rangle_{ic}\rangle$ on $<N_Q>$, that is, $\langle\langle \Delta r_i^2 \rangle_{ic}\rangle (\langle N_Q \rangle) \sim \exp(<N_Q>)$. To measure correlation, we use Pearson correlation coefficient $K$, which calculates a linear correlation coefficient of values.

$$K = \frac{E\{[X-E(X)][Y-E(Y)]\}}{D(X)D(Y)} \qquad (2)$$

where $X$ and $Y$ are two variables, $E(X)$ and $E(Y)$ are their average values, and $D(X)$ and $D(Y)$ are their standard deviations. Maximum correlation yields $K = 1$ or $K = -1$, whereas in case of no correlation one has $K = 0$. Figure 3(b) displays the Pearson correlation coefficient $K$ between $<N_Q>$ and $\ln(\langle\langle \Delta r_i^2 \rangle_{ic}\rangle)$ as a function of $n_A$. It can be seen that $K$ is between 0.2 and 0.3 for both Cu and Zr when $n_A$ equals to 1. This fact suggests that the one-to-one correspondence is rather weak. However, when $n_A$ increases, $K$ increases quickly. When $n_A$ reaches 4~10, $K$ is between 0.4 and 0.6, indicating the medium correlation. when $n_A$ is 10~40, the correlation becomes strong as K reaches 0.6~0.8. When $n_A$ is larger than 40, $K$ is larger than 0.8 and even nearly 1, suggesting the extremely strong correlation. All these facts indicate that $N_Q$ and $\langle \Delta r_i^2 \rangle_{ic}$ are closely correlated.

To reflect the direct correlation between the spatial distributions of $N_Q$ and $\langle \Delta r_i^2 \rangle_{ic}$, We colored the atoms in each group ($n_A$ = 20) with the value of $<N_Q>$ and $\langle\langle \Delta r_i^2 \rangle_{ic}\rangle$ of the group and the contoured maps of $<N_Q>$ and $\langle\langle \Delta r_i^2 \rangle_{ic}\rangle$ for all of the (Cu and Zr) atoms are shown in Fig. 4. Figure (a) and (b), (c) and (d) as well as (e) and (f) correspond to the same configuration. Figure (a), (c) and (e) show the spatial distribution of $\langle\langle \Delta r_i^2 \rangle_{ic}\rangle$ while Figure (b), (d) and (f) display the spatial distribution of $<N_Q>$. We notice that most of the atoms with higher $\langle\langle \Delta r_i^2 \rangle_{ic}\rangle$ are located in the region with high $<N_Q>$ while the distribution of the atoms with lower $\langle\langle \Delta r_i^2 \rangle_{ic}\rangle$ almost always overlaps the region with lower $<N_Q>$. This observation directly reflects the close spatial correlation between structural heterogeneity and dynamic heterogeneity.

Although close spatial correlation between $<N_Q>$ and $\langle\langle\Delta r_i^2\rangle_{ic}\rangle$ is displayed in Figure 4, the one-to-one correlation between $N_Q$ and $\langle\Delta r_i^2\rangle_{ic}$ is rather weak. Based on $N_Q$, we propose a new parameter, $\theta$:

$$\theta = CN/(CN + N_Q) \quad (3)$$

where CN is the coordination number of the atom. If $N_Q$ equals to 0, $\theta$ is 1, indicating the close atomic packing. To better reflect the correlation between local structure and dynamic heterogeneity, we introduce a combined parameter, φ:

$$\varphi = (1-x)*\theta + x*d_5 \ (0 \leq x \leq 1) \quad (4)$$

where $d_5 = n_5/CN$ [29]. $n_5$ is one of parameters in the Voronoi polyhedral index, which is expressed as $<n_3, n_4, n_5, n_6>$, where $n_i$ denotes the number of $i$-edged faces of the Voronoi polyhedron. CN is the coordination number of the atom, $CN = \sum_i n_i$. Pearson correlation coefficient $K$ as a function of $x$ between φ and $\langle\Delta r_i^2\rangle_{ic}$ are shown in Figure 5. $K(x=0)$ is larger than $K(x=1)$, indicating that the correlation between $\theta$ and $\langle\Delta r_i^2\rangle_{ic}$ is stronger than that between $d_5$ and $\langle\Delta r_i^2\rangle_{ic}$. $K(x=0.27)$ for Cu and $K(x=0.18)$ for Zr reaches the negative maximum. This fact suggests that the combined parameter $\varphi$ can better reflect the one-to-one correlation between local structure and dynamic heterogeneity.

Recently, Hocky *et al* pointed that the connection between local structure and dynamical heterogeneity in supercooled liquids is highly system dependent [30]. In different systems, the local structural metrics which can well reflect the connection might be also different. The parameter of QNA can reflect the local atomic packing. Thus we think that it can well reflect the correlation between local structure and dynamical heterogeneity in the systems with non-directional bonding, such as metallic systems, in which the atomic packing tends to be close in the solid state. It is known that dynamic propensity is determined by atomic structure, including short-range (local) and medium-range. Thus, the one-to-one correlation between local structure and dynamic propensity should be imperfect. On the other hand, one structural metric cannot fully describe local structure. Therefore, to well reflect the correlation between local structure and dynamical heterogeneity, a combination of many local structural metrics should be used. It should be noted that the role of some local structural metrics in the correlation might be more important than others, such as the QNA in metallic systems proposed in this work.

We can also use QNA to identify the atoms near the defects such as vacancy defects and dislocations in metallic crystals. Therefore, the structural feature in metallic liquids described by QNA might be considered as the existence of the "defects" in metallic liquids. Similarly, we can also use QNA to search the "defects" in metallic glasses. It is known that the defects plays an key role to determine the properties in metallic crystals. The success of QNA to describe the "defects" in metallic liquids and glasses might be helpful to uncover the structure-property relationship in metallic liquids and glasses.

## 4. Conclusion

In summary, we investigated the correlation between local structure characterized by a new parameter (QNA) and dynamic heterogeneity in a metallic glass-forming liquid, $Cu_{64}Zr_{36}$. It is found that atoms with larger $N_Q$ have higher mobility, which might be caused by the facts that atoms with larger $N_Q$ have closer atomic packing around and higher potential energy. The atoms with different $<N_Q>$ show clear spatial heterogeneities and the more (less) $<N_Q>$ regions overlap the regions with higher (lower) dynamic propensities. Therefore, we make a clear connection to link the local structure and dynamic heterogeneity. Moreover, a combined parameter is introduced to better reflect the correlation between local structure and dynamic heterogeneity.

**Acknowledgments**

This work was supported by the National Natural Science Foundation of China (Grant Nos. 51304145, 51271162 and 51574176) and the Program for the Innovative Talents of Higher Learning Institutions of Shanxi (Grant No. 143020142-S,

2014jytrc04).

**Figure Captions**

Fig. 1 (a) The schematic of QNA in a 2-D system. Atoms *C* and *D* belong to "quasi-nearest" atoms since they are an adjacent pair of the nearest neighbors of atom *A* but not nearest neighbors of each other. Similarly, atoms *A* and *B* also belong to one pair of "quasi-nearest" atoms. The blue solid line represents the nearest correlation while the red dot line corresponds to the "quasi-nearest" correlation. (b) The distribution of $N_Q$ around Cu and Zr atoms at 1000K. (c) the distribution of potential energy for Cu atoms with different $N_Q$. (d) the distribution of potential energy for Zr atoms with different $N_Q$. Insets in (c) and (d) are the $N_Q$ dependence of the average per-atom potential energy with error bars corresponding to one standard deviation.

Fig. 2 Self-intermediate scattering functions of (a) Cu and (b) Zr with different $N_Q$ at the initial time. Insets in (a) and (b) are the $N_Q$-dependence of relaxation times and the dash line corresponds to the relaxation time for the element. 8 independent runs were averaged over to calculate SISFs. The distribution of dynamic propensities ($\langle \Delta r_i^2 \rangle_{ic}$) for (c) Cu and (d) Zr with different $N_Q$. 100 runs were performed to calculate dynamic propensity for a single initial configuration at 1000K and 8 independent configurations were used. Insets in (c) and (d) are the $N_Q$ dependence of the average dynamic propensities with error bars corresponding to one standard deviation.

Fig. 3 (a) $<N_Q>$ as a function of $\langle\langle \Delta r_i^2 \rangle_{ic}\rangle$ with $n_g = 200$. For each element, the atoms are sorted by their $\langle \Delta r_i^2 \rangle_{ic}$ from low to high. They are divided into $n_g$ groups, each containing $n_A$ atoms. For each group, the average $N_Q$ ($<N_Q>$) and $\langle \Delta r_i^2 \rangle_{ic}$ ($\langle\langle \Delta r_i^2 \rangle_{ic}\rangle$) are calculated. (b) Pearson correlation coefficient as a function of $n_A$ between $<N_Q>$ and ln ($\langle\langle \Delta r_i^2 \rangle_{ic}\rangle$). 8 independent configurations were used.

Fig.4 Spatial distribution of $<N_Q>$ and $\langle\langle \Delta r_i^2 \rangle_{ic}\rangle$. We colored the atoms in each group ($n_A = 20$) with the value of $<N_Q>$ and $\langle\langle \Delta r_i^2 \rangle_{ic}\rangle$ of the group. (a) and (b), (c) and (d) as well as (e) and (f) correspond to the same configuration. (a), (c) and (e) show the spatial distribution of $\langle\langle \Delta r_i^2 \rangle_{ic}\rangle$ while (b), (d) and (f) display the spatial distribution of $\langle N_Q \rangle$.

Fig.5 Pearson correlation coefficient as a function of $x$ between a combined parameter $\varphi$ and $\langle \Delta r_i^2 \rangle_{ic}$. $\varphi = (1-x)*\theta + x*d_5$. where $\theta = CN/(CN + N_Q)$ and $d_5 = n_5/CN$. $n_5$ is one of parameters in the Voronoi polyhedral index, which is expressed as $\langle n_3, n_4, n_5, n_6 \rangle$, where $n_i$ denotes the number of $i$-edged faces of the Voronoi polyhedron. CN is the coordination number of the atoms. $CN = \sum_i n_i$. 8 independent configurations were used and the error bars correspond to one standard error.

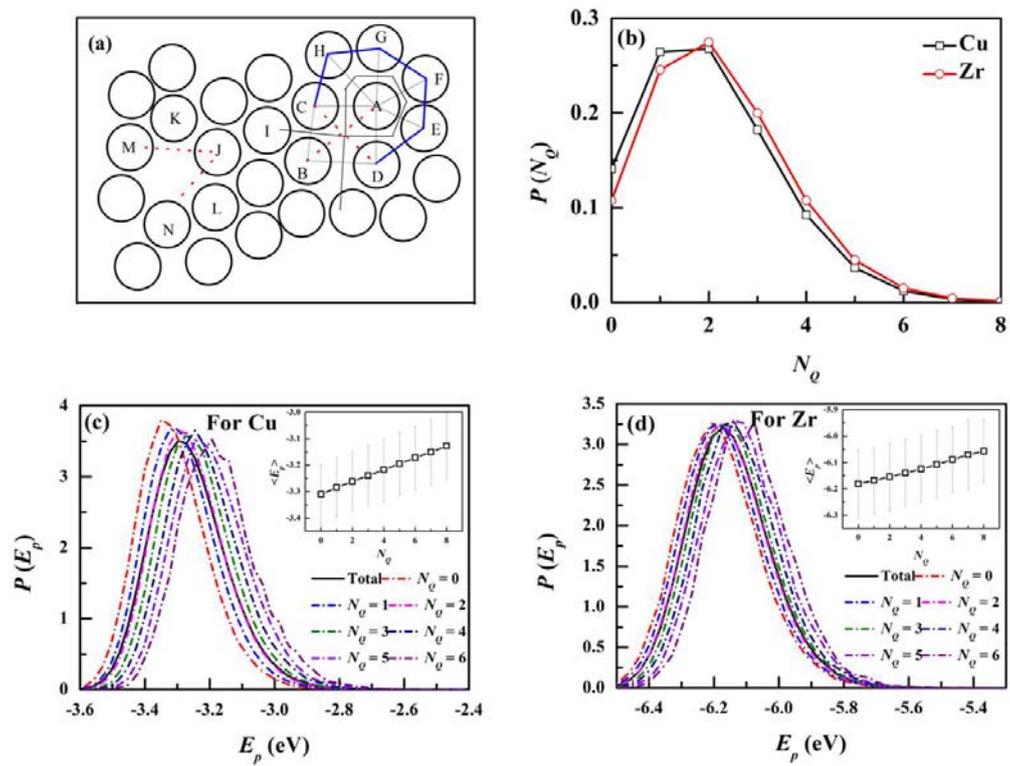

Figure 1, Pan *et al.*

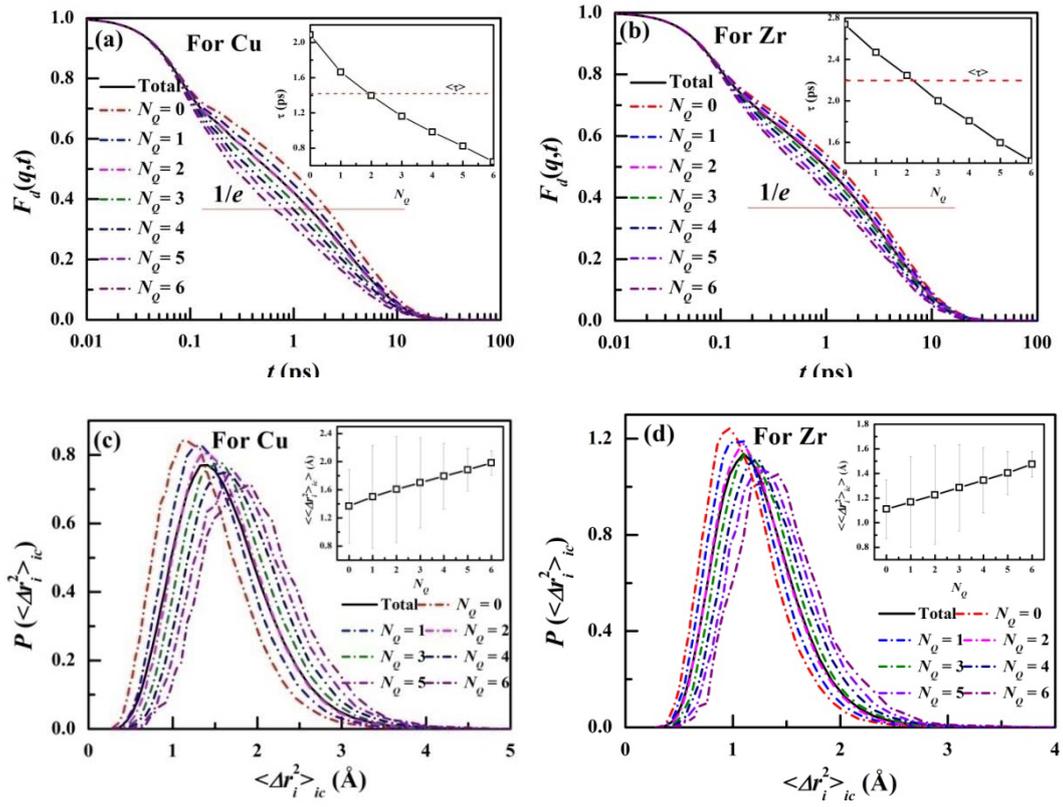

Figure 2, Pan *et al.*

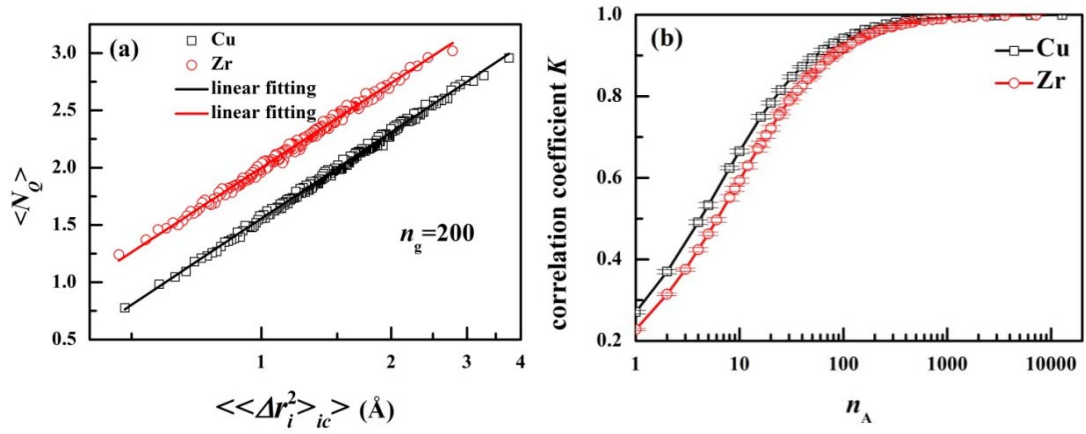

Figure 3, Pan *et al.*

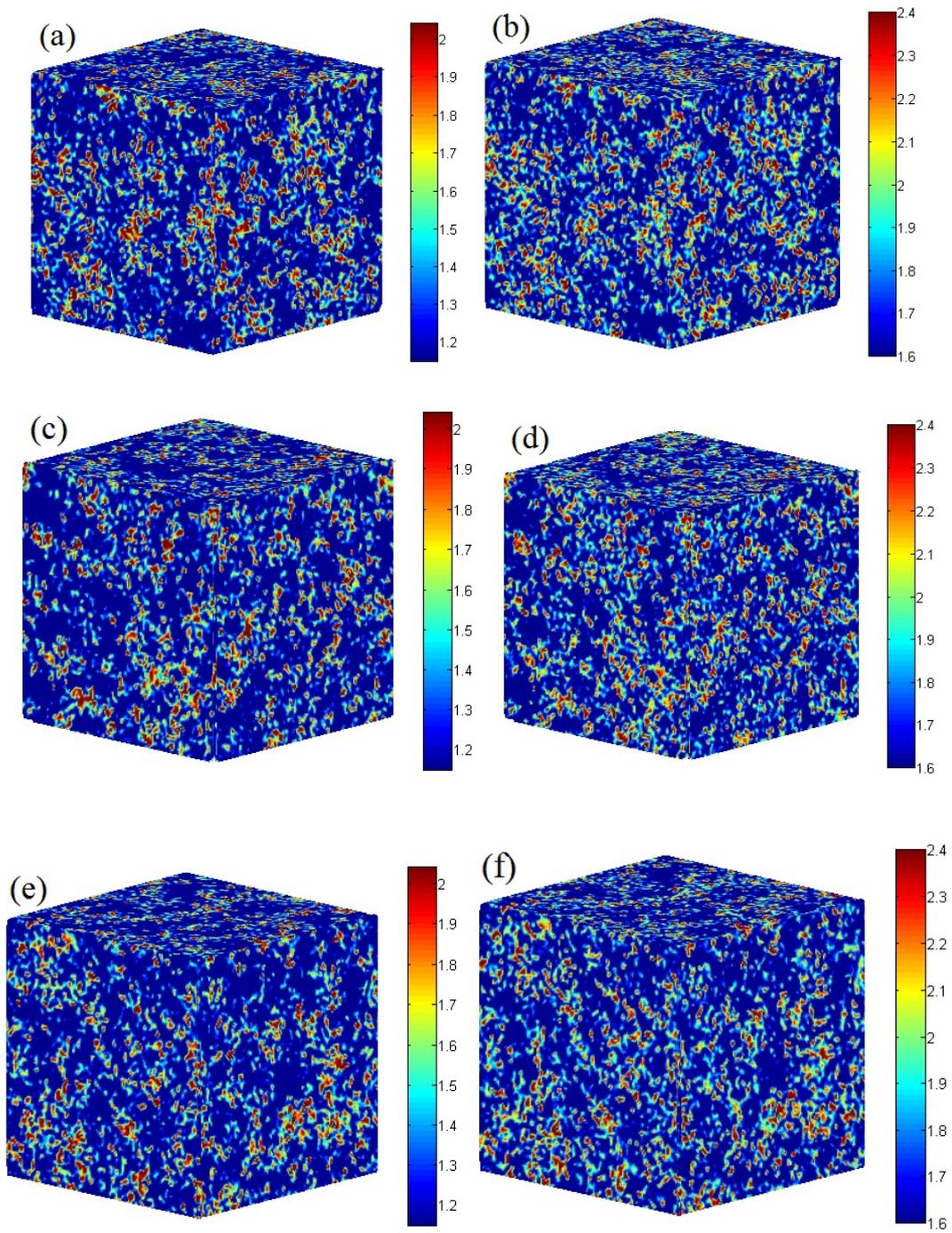

Figure 4, Pan *et al*.

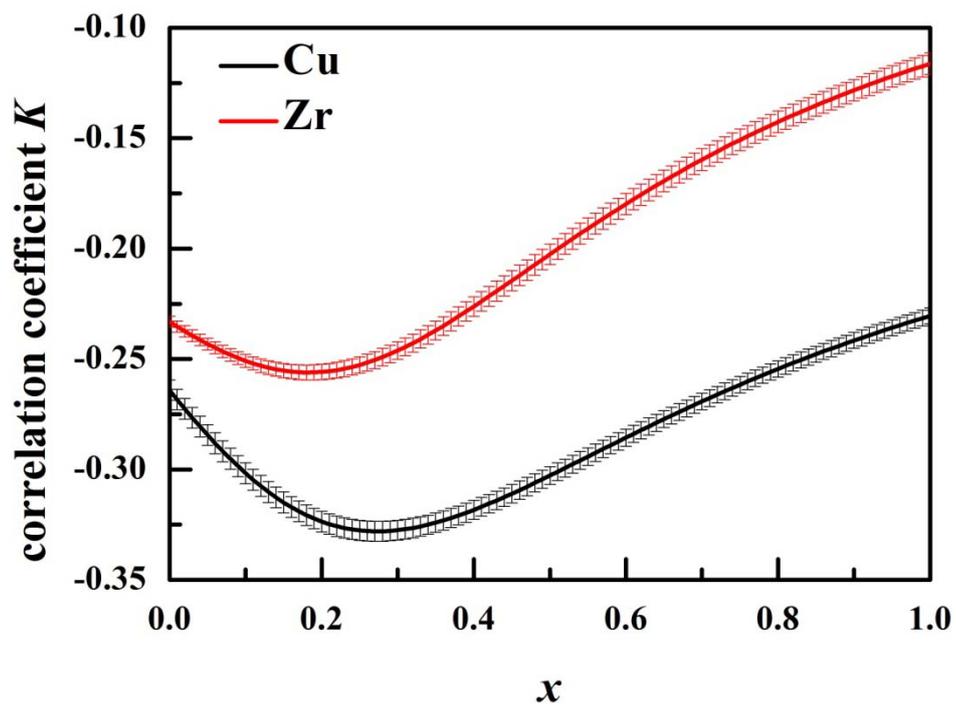

Figure 5, Pan *et al.*